# An annotated English translation of 'Kinetics of stationary reactions' [M. I. Temkin, *Dolk. Akad. Nauk SSSR*. 152, 156 (1963)]

Vladislav V. Levchenko[*], Ronan Fleming[†], Hong Qian[#], and Daniel A. Beard[*]

[*]Department of Physiology, Medical College of Wisconsin, Milwaukee, WI
[†]Science Institute & Center for Systems Biology, University of Iceland, Reykjavik, Iceland
[#]Department of Applied Mathematics, University of Washington, Seattle, WA

Address correspondence to: D. A. Beard, dbeard@mcw.edu

**Abstract**

Temkin's 1963 article on one-way fluxes and flux ratios in steady-state reaction systems bears directly on current research in physical and biological chemistry, such as in the interpretation of metabolic exchange fluxes determined from isotopomer labeling experiments. Yet, originally published in Russian [*Dolk. Akad. Nauk SSSR* 152, 156-159 (1963)], this article has remained inaccessible to much of the scientific community. Here we provide an English translation of the original article with several additional clarifications and corrections.

Temkin [*Dolk. Akad. Nauk SSSR* 152, 156-159 (1963)] derived a relationship between thermodynamic driving forces and one-way fluxes in chemical reactions that bears directly on current research in physical and biological chemistry. Indeed, material in a widely cited English-language review [M.I. Temkin, *Advances in Catalysis* 28, 173-291 (1979)] relies on primary literature, by the same author, published in Russian. This translation is intended to bring into focus the derivation of the primary result referenced in subsequent English-language literature. A complete translation of the original article follows, with additional clarifications and corrections incorporated as footnotes.

## Kinetics of stationary reactions

M. I. Temkin, *Dolk. Akad. Nauk SSSR*. **152**, 156 (1963)

Complex reactions are combinations of simple ones, called here stages or elementary reactions. Here we use the term "intermediates" to refer to chemical species that appear in the chemical equations of elementary reactions, but not in the overall chemical equation of a given complex reaction. The reaction is called stationary if the concentrations of intermediate components remain constant when the concentrations of substrates and products that appear in the overall reaction are held at fixed concentrations[1]. Such reaction regimes may be realized in flow systems or in reactors without gradients [1]. Reactions in static systems can often be considered quasi-stationary (Bodenshtein's method). Such conditions are outlined in Frank-Kamenetskii [2] and Semenov [3]. Further review is related to stationary (or quasi-stationary) reactions. In the case of heterogeneous reactions catalyzed on surfaces, we will assume the reaction surface to be homogeneous. Intermediate components for such reactions are hemosorbic[2] atoms or molecules and free surface area.

---

[1] Thus Temkin is specifically referring here to non-oscillatory steady-state systems. In general reaction systems with overall reactant concentrations held constant can display non-oscillatory or oscillatory steady states, or chaotic behavior. The study of oscillatory reaction dynamics began with the pioneering work of Belousov and Zabotinsky. Other examples include the reversible Schnakenberg reaction scheme in Vellela and Qian [*Proc. Royal Soc. A,* to appear (2010)] and the reversible Lotka-Volterra reaction in Li *et al.* [*J. Chem. Phys*. **129**, 154505 (2008)]. In both of these cases the overall reactions involve the steady-state production of product B from substrate A.

To obtain an overall complex equation by addition of elementary reactions[3], the elementary reaction equations must be multiplied by the appropriate stoichiometric numbers [4]. In general there may be substantially different sets of these stoichiometric numbers corresponding to different routes[4] for a given overall complex reaction [4]. By "substantially different" it is meant that a set of stoichiometric numbers associated with a given mechanism cannot be obtained from another set by multiplying by a common factor (which would be equivalent to multiplication of the overall complex reaction by the same factor). Stoichiometric numbers can be fractional, negative or be equal to zero, and by order of magnitude they are comparable to 1.

Let us denote $x_{js}$ as the stoichiometric coefficient of the intermediate substrate $X_j$ in a chemical equation of elementary reaction $s$: $x_{js} > 0$, if $X_j$ is formed, and $x_{js} < 0$, if $X_j$ is depleted. The stoichiometric number associated with elementary reaction $s$ for the route $p$ is denoted $v_s^{(p)}$; these numbers, by definition, satisfy

$$\sum_{s=1}^{S} v_s^{(p)} x_{js} = 0, \qquad (1)$$

where $S$ is the total number of elementary reactions. Thus there exist $I$ linearly independent conditions of the form of Equation (1), where $I$ is the number of intermediate substrates. We denote the number of linearly independent of sets of $v_s^{(p)}$—solutions of systems of the form of Equation (1) as $P = S - I$. I.e., there are $P$ independent reaction routes [4].

Sometimes there is more than one route associated with a given overall chemical equation. For example, let us assign numbers 1 and 2 to two routes that yield the same overall reaction. Then chemical equation of the route 3, defined with the stoichiometry $v_s^{(3)} = v_s^{(2)} - v_s^{(1)}$, will have the overall stoichiometry 0 = 0. We will call such routes empty. (Christiansen calls them as cyclical sequences [5].) More often there is only one independent route associated with a given overall reaction. In terms of kinetics such systems undergo one complex reaction; yet some elementary reactions can be common to different routes.

The chemical equation of an elementary reaction is written in accordance with the flow of its elementary operation. We define the flux of elementary reaction $s$ in the forward direction $r_s$ and in the reverse direction $r_{-s}$. Forward and reverse fluxes of an elementary reaction correspond to the number of turnovers of the reaction in each direction per unit volume per unit time, or per unit surface area per unit time. The chemical equation of an elementary reaction may be multiplied by an arbitrary factor (multiplier) that must be specified[5]. Given these definitions, the turnover of a reaction is associated with the disappearance of molecules of initial substances (substrates) and appearance of molecules of products in a number that is indicated in a coefficient of the chemical equation. Denoting the rate of a given reaction route $p$ as $r^{(p)}$ (corresponding to rate of turnover of a given reaction route in a given time period per unit volume or per unit surface area), the total flux through a given elementary reaction is the sum of contributions from the $P$ routes, the following criteria of stage stationarity are satisfied:

$$\sum_{p=1}^{P} v_s^{(p)} r^{(p)} = r_s - r_{-s}.  \qquad (2)$$

---

[2] I.e., the reaction surface is uniform; there is no spatial heterogeneity in the surface absorption or reaction kinetics.

[3] We apply the term "elementary reaction" where Temkin uses the term "стадия", which may be more properly translated as "stage". We use the more conventional contemporary notation "elementary reaction" throughout.

[4] The term "route" appears spelled in Latin characters in the original paper. Possible alternative translations are "pathway" or "mechanism". We use the term "route" throughout, which refers to a set of elementary reactions and associated stoichiometric numbers for each elementary reaction.

And so indeed, in a given volume or on a given surface, $\sum_{s=1}^{S} x_{js}(r_s - r_{-s})$ molecules of $X_j$ are formed per unit time. Combining Equations (1) and (2), we have

$$\sum_{s=1}^{S} x_{js}(r_s - r_{-s}) = 0; \qquad (3)$$

i.e. number of each intermediate substance is constant. For a given reaction system, a total of $S$ equations from Equation (2) may be used to define $P + I$ unknowns. That is, $P$ values of $r^{(p)}$ and $I$ values of activity of independent intermediate substances. Equation (3) may be obtained as a direct consequence of stationary-state mass conservation.

A transformation from one set of independent routes $\{v_s^{(1)}, v_s^{(2)}, \ldots\}$ to another $\{v_s^{(1')}, v_s^{(2'')}, \ldots\}$ may be defined by equations of the form

$$\begin{aligned} v_s^{(1')} &= C_{11} v_s^{(1)} + C_{12} v_s^{(2)} + \ldots \\ v_s^{(2')} &= C_{21} v_3^{(1)} + C_{22} v_s^{(2)} + \ldots, \\ &\vdots \end{aligned} \qquad (4)$$

where $C_{11}$, $C_{12}$, etc. are coefficients. Therefore the chemical equations of the route are transformed. If $\{r^{(1)}, r^{(2)}, \ldots\}$ are the route reaction rates for the original set of routes and $\{r^{(1')}, r^{(2')}, \ldots\}$ are the rates for the new set of routes, then

$$\begin{aligned} r^{(1)} &= C_{11} r^{(1')} + C_{21} r^{(2')} + \ldots \\ r^{(2)} &= C_{12} r^{(1')} + C_{22} r^{(2')} + \ldots \\ &\vdots \end{aligned} \qquad (5)$$

Coefficients along a row in the system of Equation (4) appear in the corresponding column in the system of Equation (5)[6]. Indeed, substituting in the reaction route rates from Equation (5) into Equation (2) and taking into account Equation (4), we find that the route rates $\{r^{(1')}, r^{(2')}, \ldots\}$ satisfy the conditions of stationarity for all $s$. It is worth noting that the left-hand side of Equation (2) has a form of a scalar product of two vectors. The steady-state operation of chemical transformations may be examined using linear algebra when routes are decomposed into elementary reactions.

---

[5] Multiplication of a chemical reaction by an arbitrary factor does not change the chemistry, but does change the law of mass action. Therefore the factor has to be defined and is not arbitrary.

Let us assume the routes 1 and 2 correspond to the same overall chemical equation. Defining the transformation $v_s^{(1')} = v_s^{(1)}$, $v_s^{(2')} = v_s^{(2)} - v_s^{(1)}$ (where $v_s^{(2')}$ defines an empty route), it follows from Equation (5) that $r^{(1')} = r^{(1)} + r^{(2)}$ and $r^{(2')} = r^{(2)}$. Therefore the rate of reaction on an empty route is not in general equal to zero. (But it is always equal to zero if the empty route is the only route). Although route 1 is left unchanged by the transformation, $r^{(1')} \neq r^{(1)}$. Therefore it is not possible to unambiguously assign a defined rate to this route, unless the set of all other independent routes and their corresponding rates are defined.

Using these relationships it is possible to define a unique "summary" route for which the reaction rate of all other routes is equal to zero. The stoichiometric coefficient for elementary reaction $s$ in the summary route is defined by the flux-weighted sum of coefficients $v_s^{(p)}$ from an independent set of routes:

$$v_s = \frac{\sum_{p=1}^{P} v_s^{(p)} r^{(p)}}{\sum_{p=1}^{P} r^{(p)}} = \frac{1}{r} \sum_{p=1}^{P} v_s^{(p)} r^{(p)}. \tag{6}$$

The reaction rate associated with the route defined by this set of stoichiometric coefficients $\{v_s\}$ (the overall reaction flux) is equal to $r = \sum_{p=1}^{P} r^{(p)}$. The flux through each other reaction route in the new transformed set of routes is zero. The chemical equation of the summary route describes the total (sum of all) chemical transformation in the system.

In some cases the elucidation of kinetic equations relating route reaction fluxes $\{r^{(1)}, r^{(2)}, \ldots\}$ to concentrations (or activities) of substrates and products can be significantly simplified this way. Starting with the equality

$$(r_1 - r_{-1}) r_2 r_3 \ldots + r_{-1} (r_2 - r_{-2}) r_3 \ldots + r_{-1} r_{-2} (r_3 - r_{-3}) \ldots + \ldots = r_1 r_2 r_3 \ldots - r_{-1} r_{-2} r_{-3} \ldots \tag{7}$$

take $(r_1 - r_{-1})$, $(r_2 - r_{-2})$, and so on according to (2). Dividing both sides of equality by $r_1 r_2 r_3 \ldots$ we get

---

[6] The relation between Equations (4) and (5) can be demonstrated as follows. Expressing Equation (4): $v_s^{(p')} = \sum_{p=1}^{P} C_{p'p} v_s^{(p)}$, if the total flux through each elementary reaction is maintained by this transformation, then $\sum_{p'=1}^{P} v_s^{(p')} r^{(p')} = r_s - r_{-s}$, which implies $r^{(p)} = \sum_{p'=1}^{P} C_{p'p} r^{(p')}$.

$$r^{(1)}\left(\frac{v_1^{(1)}}{r_1}+\frac{r_{-1}v_2^{(1)}}{r_1r_2}+\frac{r_{-1}r_{-2}v_3^{(1)}}{r_1r_2r_3}+\ldots\right)+r^{(2)}\left(\frac{v_1^{(2)}}{r_1}+\frac{r_{-1}v_2^{(2)}}{r_1r_2}+\frac{r_{-1}r_{-2}v_3^{(2)}}{r_1r_2r_3}+\ldots\right)+\ldots=1-\frac{r_{-1}r_{-2}r_{-3}\ldots}{r_1r_2r_3\ldots}. \quad (8)$$

Equation (8) may be called the equation of stationary reactions. Along with Equation (2) this equation is independent of the order in which the elementary reactions are numbered. This is obvious because all elementary reactions in a stationary reaction run simultaneously.

In some cases simplifications are possible. For example, sometimes the catalytic mechanism for an overall reaction contains a elementary reaction in which $v_s^{(p)}=0$ for all $p$. Such elementary reactions are in equilibrium ($r_s = r_{-s}$) and the factors $r_{-s}/r_s$ fall out of Equation (8). Furthermore, when $r_s$ for one of the elementary reactions is much greater than $r_s - r_{-s}$, this elementary reaction is referred to as "rapid" or "quasi-equilibrium". Elementary reactions where $r_s$ is of similar order to $r_s - r_{-s}$ are called "slow". According to Equation (2) for fast elementary reactions $\left|v_s^{(p)}r^{(p)}\right| \ll r_s$ and $r_s \approx r_{-s}$, leading to associated simplifications of Equation (8). If some elementary reaction $s$ is treated as irreversible, $r_{-s}$ becomes zero.

It is sometimes possible to easily eliminate some unknown activities (or concentrations) of intermediate substances based on Equation (8). To do so, it is necessary if possible to arrange the equations of elementary reactions in such an order that intermediate substances which are formed in any given elementary reaction are consumed in the following elementary reaction. In some cases a particular ordering allows one to obtain the necessary quantity of equations. Diagrams that show the structure of the reaction mechanism can often be helpful [6, 3, 5]. Concentrations of intermediate substances, if need be, may be determined with the help of the corresponding elementary reaction sequences. If equilibrium or quasi-equilibrium elementary reactions are present then the activity of an intermediate substance which is participating in such a elementary reaction can be elucidated from the equilibrium mass-action ratio. Applying Equation (8) in the context of the summary route where all other routes have zero flux, we have

$$r\left(\frac{v_1}{r_1}+\frac{r_{-1}v_2}{r_1r_2}+\frac{r_{-1}r_{-2}v_3}{r_1r_2r_3}+\ldots\right)=1-\frac{r_{-1}r_{-2}r_{-3}\ldots}{r_1r_2r_3\ldots} \quad (9)$$

An equation of this form can be applied directly to any overall reaction with only one route.

Let us next consider a summary route of the complex reaction (or equivalently the route associated with a single-route reaction) in which one of the elementary reactions is irreversible. We further assume, without loss of generality, that the elementary reactions are ordered so that the irreversible elementary reaction is elementary reaction $S$. Therefore Equation (9) yields

$$r_+\left(\frac{v_1}{r_1}+\frac{r_{-1}v_2}{r_1r_2}+\frac{r_{-1}r_{-2}v_3}{r_1r_2r_3}+\ldots\frac{r_{-1}r_{-2}\ldots v_S}{r_1r_2\ldots r_S}\right)=1 \quad (10)$$

where $r_+$ is the forward flux associated with elementary reaction $S$.

If only one elementary reaction is slow (others are quasi-equilibrium or equilibrium), this elementary reaction is called limiting. For such a case we get equality $r_+ = r_l/v_l$ from (10), where $l$ is the index of the limiting elementary reaction[7]. By changing the direction of all elementary reactions and aligning them in the reverse order and equaling $r_l = 0$, we derive the equation for the rate in the reverse direction, $r_-$. Then it is simple to see that[8]

$$\frac{r_-}{r_+} = \frac{r_{-1} r_{-2} r_{-3} \cdots}{r_1 r_2 r_3 \cdots};  \quad (11)$$

$$r = r_+ - r_- \quad (12)$$

According to Equations (11) and (12) the values of $r_+$ and $r_-$ are independent of the order in which the elementary reactions are aligned (because $r_+/r_-$ and $r_+ - r_-$ do not change when the order of the elementary reactions changes).

When we apply Equations (8)-(12) suggested by Christiansen (5,7) to sequences of elementary reactions of special type, we arrive to the results suggested by this author.

When we have only one way of isotopic exchange then the rate of the isotopic exchange is described by an equation of the same form as Equation (10), but includes only stages that are participating in the exchange and not all elementary reactions of the overall reaction. Such an equation was derived by Matsuda and Horiuti [8].

Let us take the reaction with one route, with the chemical equation of aA + a'A' + … = bB + b'B' + … We define the average (mean) stoichiometric value $\bar{v}$:

$$\bar{v} = \frac{v_1 \Delta G_1 + v_2 \Delta G_2 + \cdots}{\Delta G_1 + \Delta G_2 + \cdots}. \quad (13)$$

Here $\Delta G_s$ is the change of Gibbs free energy associated with elementary reaction $s$. From the theory of absolute rates of reactions it is evident that delta $\Delta G_s = -RT \ln \frac{r_s}{r_{-s}}$, as well as

$$v_1 \Delta G_1 + v_2 \Delta G_2 + \ldots = \Delta G = -RT \ln K \frac{(A)^a (A')^{a'} \ldots}{(B)^b (B')^{b'} \ldots},$$ where $K$ is equilibrium constant, $(A)$ is the activity of substance (substrate) A and so on. It follows that Equation (11) yields:

$$\frac{r_+}{r_-} = \left[ K \frac{(A)^a (A')^{a'} \ldots}{(B)^b (B')^{b'} \ldots} \right]^{1/\bar{v}}. \qquad (14)$$

---

[7] There is some confusion here, perhaps in part due to translation. In the paragraph before Equation (10) the limiting reaction is defined to be the "last" elementary reaction. We have taken the liberty of clarifying that the index of the last elementary reaction is defined to be $S$ above. Thus the equality $r_+ = r_l/v_l$ is true only if $l = S$.

[8] Several issues related to Equations (11) and (12) require clarification. First, note that the $r_+$ and $r_-$ obtained this way do not correspond to the overall forward and reverse rates for the overall reaction. Indeed, this incorrect assumption was made in a similar proof described by T.L. Hill [*Free Energy Transduction in Biology*, Academic Press Inc., U.S. (1977) p. 22] and was later demonstrated to be incorrect by Hill [cf. T.L. Hill, *Free Energy Transduction and Biochemical Cycle Kinetics*, Dover, (2004) p. 53]. Second, while the analysis here is limited to systems in which all but one elementary reaction are in quasi-equilibrium, the relationships of Equations (11) and (12) do hold true for systems of this sort for the more general case in which any number of the reactions are limiting, as was demonstrated correctly by Hill [*Free Energy Transduction and Biochemical Cycle Kinetics*, Dover (2004) and references therein], and Qian *et al.* [*Scientia Sinica* **24**, 1431-1448 (1981); **25,** 31-40 (1982); **27**, 470-481 (1984)]. In a more recent work, Beard and Qian [*PLoS ONE* 2: e144, 2007] provide a simpler proof of a more general theorem that applies to a broader set of reaction and transport processes than are considered here.

In the Equation (13) contributions related to equilibrium or quasi-equilibrium elementary reactions drop out because delta $\Delta G_s = 0$ for them (exact or approximate). If $v_s$ of all non-equilibrium elementary reactions are the same, then $\bar{v}$ is equal to that overall value[9]. In particular, if we have limiting elementary reactions, then $\bar{v} = v_l$ and we arrive to the results of Boreskov [9] and Horiuti [4].

---

[9] Recall that, here, Equation (11) was proved (albeit incorrectly) strictly for the case where there is only one non-equilibrium elementary reaction. Therefore there is only one non-equilibrium elementary reaction.